\documentclass[prb, twocolumn,amsmath,amssymb, superscriptaddress,longbibliography, nofootinbib,longbibliography]{revtex4-2}
\usepackage[pdftex,colorlinks=true]{hyperref}
\hypersetup{
citecolor = black
}
\usepackage{graphicx}% Include figure files
\usepackage{natbib,hyperref}
\usepackage{amsmath,epsfig,color,amssymb}
\usepackage{bbold}
\usepackage[dvipsnames]{xcolor}
\usepackage[export]{adjustbox}
\usepackage{comment}
\newcommand{\beg}{\begin{equation}}
\newcommand{\en}{\end{equation}}

\newcommand \bel  {\begin{align}}
\newcommand \enl  {\end{align}}

\newcommand{\dg}{^\dagger}
\newcommand{\ket}[1]{|#1\rangle}
\newcommand{\bra}[1]{\langle#1|}

\definecolor{new}{rgb}{.08,.05,.8}

\begin{document}

\author{Ammar Kirmani}
\affiliation{Physics Department, City College of the City University of New York, NY 10031, U.S.A.}

\author{Derek S. Wang}
\affiliation{IBM Quantum, IBM T.J. Watson Research Center, Yorktown Heights, 10598, U.S.A.}

\author{Pouyan Ghaemi}
\affiliation{Physics Department, City College of the City University of New York, NY 10031, U.S.A.}
\affiliation{Physics Program, Graduate Center of City University of New York, NY 10031, U.S.A.}

\author{Armin Rahmani}
\affiliation{Department of Physics and Astronomy and Advanced Materials Science and Engineering Center,
Western Washington University, Bellingham, Washington 98225, U.S.A.}

\title{Braiding fractional quantum Hall quasiholes on a superconducting quantum processor}

\begin{abstract}
Direct experimental detection of anyonic exchange statistics in fractional quantum Hall systems by braiding the excitations and measuring the wave-function phase is an enormous challenge. Here, we use a small, noisy quantum computer to emulate direct braiding within the framework of a simplified model applicable to a thin cylinder geometry and measure the topological phase. Our algorithm first prepares the ground state with two quasiholes. It then applies a unitary operation controlled by an ancilla, corresponding to a sequence of adiabatic evolutions that take one quasihole around the other. We finally extract the phase of the wave function from measuring the ancilla with a compound error mitigation strategy. Our results open a new avenue for studying braiding statistics in fractional Hall states.

\end{abstract}

\maketitle
\section{Introduction}

Fractional quantum Hall (FQH) states are correlated states of matter in which topological order emerges from strong electron-electron interactions. They have unique properties, such as excitations with fractional charge and fractional statistics \cite{Wilczek1984}. Experimental confirmation of the latter is especially challenging: only recently have signatures of fractional statistics been observed through interferometry and noise signals probed by transport measurements in FQH systems with filling fraction $\nu=\frac{1}{3}$ \cite{Nakamura2020,Bartolomei2020}. In addition to noise probes \cite{Lee2022}, there have been theoretical proposals for detecting signatures of anyonic braiding in the quench dynamics \cite{quenchbraiding}. Still, detecting the anyonic statistics of FQH states by directly braiding one excitation around the other through adiabatic Hamiltonian evolution remains an open experimental challenge. Meanwhile, there have been extensive advancements in theoretical models for the fractional Hall states, such as one-dimensional (1D) model Hamiltonians \cite{Seidel2005,PhysRevLett.97.056804,Feiguin2008,SiedelX,Nakamura13} that are particularly suitable for implementation on small, noisy quantum devices \cite{PRXQuantum.1.020309,ourPRL}, that capture two-dimensional (2D) physics.
Such devices have been used for studying topological braiding in the toric and surface codes without emulating Hamiltonian evolution~\cite{googleb, PRXQuantum.3.040315}. Furthermore, the braiding of Majorana fermions on quantum computers has been studied in Refs. \cite{Pekker21,Movassagh}. 

In this paper, we realize fractional statistics of FQH states by emulating direct adiabatic braiding of FQH quasiholes on IBM's digital quantum computers with a composite error mitigation strategy. Our results demonstrate the potential of small, noisy quantum devices to realize novel quantum phenomena that are challenging to study in natural systems, especially the dynamical properties of correlated electron phases.

The remainder of the paper is organized as follows. In Sec. \ref{sec:model}, we present a 1D model that captures the physics of the 2D FQH system in cylindrical geometry for a range of cylinder circumferences. We then discuss the quasihole states within the framework of this model. Then, in Sec.~\ref{sec:parent}, we discuss the parent Hamiltonian and the adiabatic braiding process. In Sec. \ref{sec:algorithm}, we discuss the quantum algorithm for the implementation of braiding and measurement of the statistical Berry phase. The results of executing the algorithm on IBM quantum hardware are presented in Sec.~\ref{sec:ibm}. Finally, the conclusions are presented in Sec.~\ref{sec:conclusions}. The appendices include details of the interpolation scheme for approximating the wave function along two segments of the trajectory, as well as examples of the circuits for preparation of the initial state and implementation of the braiding unitaries.

\section{1D model for quasiholes}\label{sec:model}
When a 2D electron gas is placed in a perpendicular magnetic field, the dispersing electronic band structure transforms into a set of degenerate  levels separated by a finite gap. The FQH states result from the interaction between electrons partially filling one Landau level \cite{fqhscience,RevModPhys.71.S298}. The effective Hamiltonian for FQH states corresponds to the projection of the {electron-electron} interaction into the degenerate orbitals of a partially filled Landau level. In this paper, we focus on fractional quantum Hall states in the cylinder geometry \cite{Siedel13}. We set the magnetic length to 1 and use the Landau gauge with the vector potential in the $x$ direction. The Landau orbital wave functions are then given by $\phi_n=\frac{1}{\sqrt{\pi^{1/2}L_x}}e^{-in\kappa x-\frac{(y-n\kappa)^2}{2}}$,
where $\kappa=2\pi/L_x$. The many-body state in the lowest Landau level is labeled by the occupation number of the degenerate Landau orbitals. The effective Hamiltonian with interactions projected onto the lowest Landau level is then given by
\beg\label{hfullapp}
\hat{H}=\sum_{j=0}^{N_\phi -1}\sum_{k>|m|}V_{k,m}c_{j+m}\dg c_{j+k}\dg c_{j+m+k}c_j,
\en
where the coefficients $V_{k,m}$ supporting a Laughlin-type state are
\beg
V_{k,m}\propto(k^2-m^2)\exp\left(-\kappa^2\frac{(k^2+m^2)}{2 }\right),
\en
with $N_{\phi}$ representing the number of flux quanta, or the degeneracy of the Landau level~\cite{Nakamura13}. Each of the $N_{\phi}$ Landau orbitals corresponds to a site in 1D model. Although interactions lead to coupling between electrons in all orbitals, in the case of thinner cylinders or large $\kappa$, longer-range interactions are suppressed. The interaction involves two-particle hopping by $m$ lattice sites in opposite directions, preserving the center of mass. As a result, the momentum operator $\hat{K}=\sum_{j}jc_j\dg c_j$ is conserved. Thus, the many-body eigenstates of the Hamiltonian in Eq.~\ref{hfullapp} can be labeled by its eigenvalue $K$. 

In the 1D limit where $L_x\to 0$, the ground state of Hamiltonian (\ref{hfullapp}) is the charged-density-wave (CDW) state $\ket{100100...100}$. For fractional-$1/3$ systems, there are three CDW states $\ket{100100...}$, $\ket{010010...}$ and $\ket{001001...}$, which correspond to three topological sectors labeled by $c=0,1,2$, respectively. A unitary operator $\hat{S}(L_x,L_x')$ transforms the ground state with circumference $L_x'$ into the ground state with circumference $L_x$ \cite{Siedel13}. The Laughlin-type state in one topological sector is then given by 
\begin{equation}\label{root2gs}
    \ket{\Psi_{1/3}(L_x)}=\hat{S}(L_x,0)\ket{100100..},
\end{equation}
with the CDW state serving as a root~\cite{Nakamura13}. The approximate form of $\hat{S}(L_x,0)$ is given in Ref.~\cite{Nakamura13} for $L_x\leq 7\ell_B$ and was subsequently implemented on a quantum computer \cite{PRXQuantum.1.020309} to generate the state in Eq. (\ref{root2gs}) up to $7\ell_B$. Using this truncated model, the FQH geometric excitations were simulated on IBM quantum devices \cite{ourPRL}. 

For quasihole states, we note that inserting a 0 in the CDW $\ket{100100100...}\to \ket{100100\underline{0}100}$ corresponds to adding a flux quantum to the system while keeping the number of electrons $N$ fixed. The state $\hat{S}(L_x,0)\ket{100100\underline{0}100...}$ is then a quasihole eigenstate of Hamiltonian in Eq. (\ref{hfullapp}) \cite{JoliEdge}. 
Introducing the reduced register notation of Ref. \cite{PRXQuantum.1.020309},
\[
\ket{\underbrace{100}_{q=0}\underbrace{100}_{q=1}\underbrace{100}_{q=2}...\underbrace{100}_{q=N-1}},
\]
where each block of three consecutive sites is represented by a reduced register $q$, we label the CDW patterns where the empty site 0 is inserted between the $q$th and $(q+1)$th reduced register as $\ket{c,q}$, e.g., $\ket{0,1}=\ket{100100\underline{0}...100}$. Then $\hat{S}(L_x,0)\ket{c,q}$ is the quasihole state with the additional flux quantum inserted at position $\kappa b_{c,q}$ in the $L_x=0$ root pattern, where $b_{c,q}=3q+2+c$ for the $\nu=1/3$ system \cite{Siedel13}. 
%Note that states $\ket{c,q}$ that differ in $c,q$ also differ in $K$ and thus are orthogonal.
To generate the braiding statistics, we must move a quasihole in the 2D plane. Thus, we create a state where a quasihole is localized at a point $h=(h_x,h_y)$ in two dimensions as in Ref. \cite{Siedel13}:
\beg
\begin{split}\label{hole1}
	&| \psi_c(h) \rangle={\cal N}\sum_q \phi_{c,q}(h)\hat{S}(L_x,0)|c,q\rangle,\\
	\phi_{c,q}(h)&=\mathrm{exp}\bigg(-iq(h_x\kappa+\pi)-\frac{1}{6}(h_y-\kappa b_{c,q})^2\bigg),
\end{split}
\en 
where ${\cal N}$ is the normalization constant.
 \begin{figure}[!h]
    \includegraphics[width=8cm]{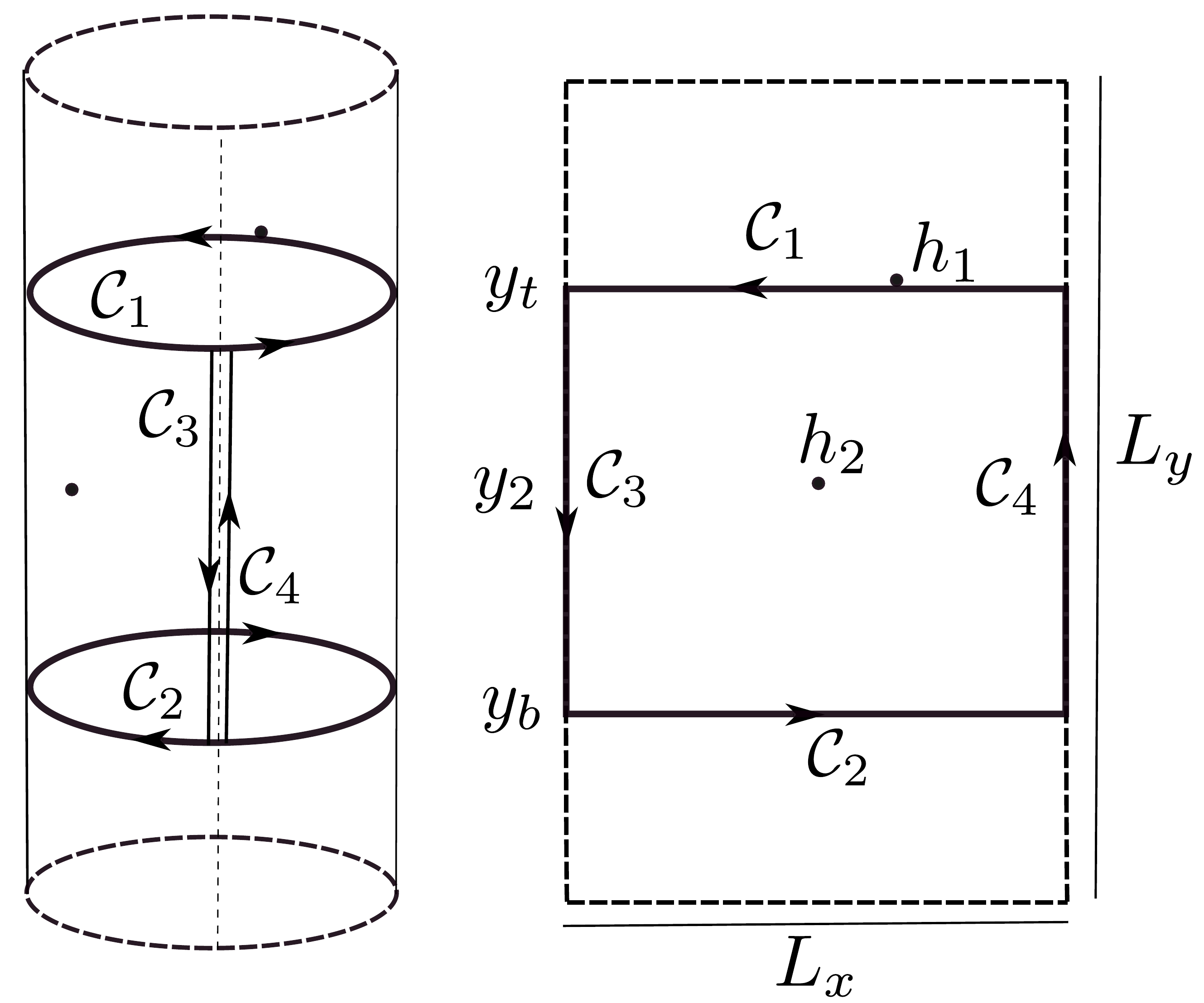}
    \caption{Schematic diagram of braiding process on a cylinder. The quasihole at $h_1$ completes a cyclic trajectory around $h_2$ by moving through two horizontal contours $\mathcal{C}_1$ and $\mathcal{C}_2$ along the circumference of the cylinder and two vertical contours $\mathcal{C}_3$ and $\mathcal{C}_4$ along the axis of the cylinder. Contours $\mathcal{C}_3$ and $\mathcal{C}_4$ run in opposite directions and do not contribute to the total Berry phase.}
    \label{fig:schematic}
\end{figure}

\begin{figure}[!h]    
	\includegraphics[width=8cm]{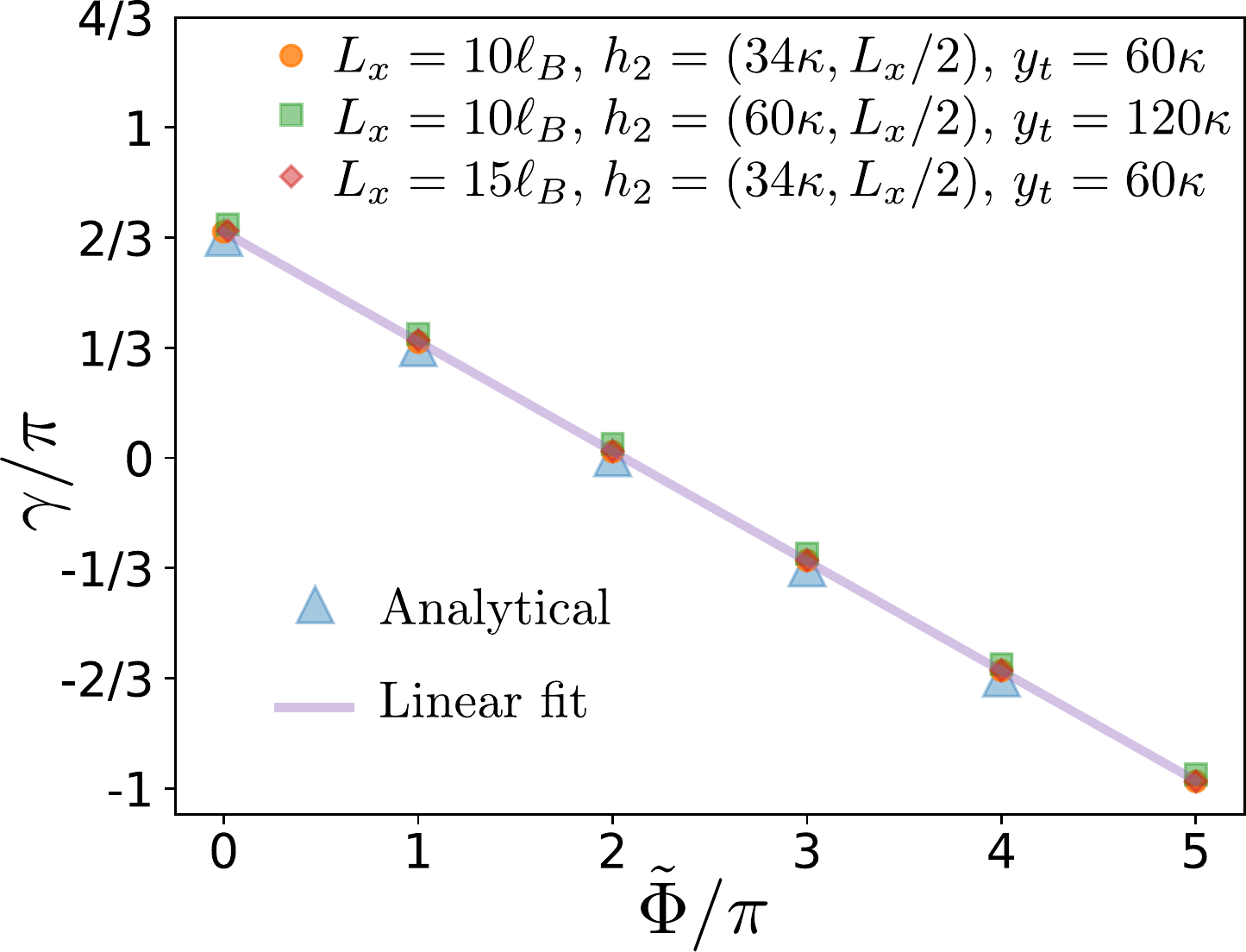}
	\centering
	\caption{The Berry phase as a function of $\tilde{\Phi}=\Phi ~{\rm mod}~ (6\pi)$, where $\Phi$ is the flux through the braiding trajectory for various parameter sets. The numerically obtained Berry phase for all paths parametrized by circumference $L_x$, quasihole position $h_2$, and $y_t$, the $y$ coordinate of ${\cal C}_1$ trajectory (see Fig.~\ref{fig:schematic}) agrees with the analytical result of Eq.~\eqref{analytic}.}
    \label{Fig:FullPath}
\end{figure}
Similarly, a two-quasihole state can be written as \cite{Siedel13}
\beg\label{hole2}
\ket{\psi_c(h_1,h_2)}=\sum_{q_1<q_2}\phi_{c,q_1,q_2}(h_1,h_2)\hat{S}(L_x,0)\ket{c,q_1,q_2},
\en
where $\ket{c,q_1,q_2}$ has two additional zeros inserted at reduced registers $q_1$ and $q_2$. As an example, $|0,2,4\rangle=\ket{100100100,\underline{0},100100,\underline{0},100...}$. When $\kappa(q_2-q_1)\gg1$, the wave function amplitude factorizes as \cite{Siedel13}
\beg\label{eq:factor}
\phi_{c,q_1,q_2}(h_1,h_2)\approx \mathcal{N}_2 \phi_{c,q_1}(h^-)\phi_{c+1,q_2}(h^+),
\en 
where $(h^-,h^+)=(h_1,h_2)$ for $h_{1y}<h_{2y}$ and $(h^-,h^+)=(h_2,h_1)$ otherwise. In the above expression, $\mathcal{N}_2$ is the normalization constant. The important shift $c\to c+1$ in the topological sector of the second quasihole is due to the presence of the first quasihole.

\section{Adiabatic braiding and parent Hamiltonian}\label{sec:parent}
We are interested in the braiding statistics when one quasihole fully encircles the other, as shown in Fig \ref{fig:schematic}. When one quasihole is adiabatically moved around the other, there are two contributions to Berry phase $\gamma$: one arises from the Aharonov-Bohm effect and is proportional to the magnetic flux $\Phi$ enclosed by the encircling path, and the other is from the anyonic exchange phase $2\pi/3$ of two quasiholes \cite{Siedel13}: 
\beg\label{analytic}
\gamma=-i\oint d\vec{l}. \bra{\psi_c(h_1,h_2)} \nabla_{h_1} \ket{\psi_c(h_1,h_2)}=-\frac{\Phi}{3}+\frac{2\pi}{3}.
\en
We consider the braiding trajectory shown in Fig. \ref{fig:schematic}. Two comments are in order regarding this setup. First, the flux through the trajectory is given by $\Phi= L_x (y_t-y_b)$. Therefore a shift of $3\kappa$ to $y_t-y_b$ changes the flux by $6\pi$, giving rise to an equivalent Berry phase. All fluxes are then equivalent up to shifts of $6\pi$, and we can only focus on flux modulo $6\pi$, represented by $\tilde \Phi$.
Second, the factorization of the wave function in Eq. (\ref{hole2}) is not valid unless $|h_{1y}-h_{2y}|\gg1$. 
 Notably, the wave function of Eq. (\ref{hole2}) is not continuous when the $y$ components of the two quasiholes cross along the $\mathcal{C}_3$ and $\mathcal{C}_4$ paths. As discussed in Appendix \ref{app:unphys}, we approximate the wave function by a continuous function by promoting the topological sectors to continuous variables and interpolating the phase. The contributions of paths $\mathcal{C}_3$ and $\mathcal{C}_4$ to the Berry phase cancel each other, so this approximation has no effect on the measured exchange statistics. 
  Thus, the only paths that contribute to the Berry phase are $\mathcal{C}_1$ and $\mathcal{C}_2$.
 
  To obtain the correct exchange statistics, we need sufficient separation between the two quasiholes. We also need to capture the liquid nature of the phase to have a nearly uniform charge density.
 Thus, both the length $L_y$ and the circumference $L_x$ of the cylinder must be large enough. It has been found that $L_x>9\ell_B$ is sufficient to realize the liquid limit~\cite{Lee2004}. We have found that the two conditions needed to obtain the correct Berry phase can be satisfied for $N\sim {L_xL_y\over 6\pi}>24$. In this paper, we use values of $N$ between 24 and 30.
 
To emulate the braiding of the quasiholes on a small, noisy quantum device, we must change the position of the quasihole must adiabatically. The adiabatic theorem, however, is formulated for the slow parametric change of a Hamiltonian, which then results in the transformation of the wave function. Thus, we need to construct parent Hamiltonians that support the quasihole states as their eigenstates. Physically, to braid the quasiholes in a 2D electron gas, one would apply a confining potential that localizes the quasiholes and breaks the degeneracy between different quasihole positions. This operation adds a potential term to the Hamiltonian, parameterizing the position of the quasihole \cite{JoliEdge}. 

We take an alternative approach suitable for quantum computers to extract the Berry phase by efficient use of quantum resources. We can directly construct a parent Hamiltonian 
\begin{equation}\label{eq:parent}
H_{\rm pr}(h_1,h_2)=\openone-|\psi(h_1,h_2)\rangle\langle \psi(h_1,h_2)|,
\end{equation}
which supports our quasihole wave function as a zero-energy eigenstate. 
All of the time dependence is then encoded in the parameter $h_1$, the position of the moving quasihole. All other eigenvalues of $H_{\rm pr}$ are degenerate and separated from the zero-energy ground state by a gap equal to one. We can see this by noting that we can write the identity in Eq.~\eqref{eq:parent} as its resolution in terms of all eigenstates of $H_{\rm pr}$ as $\openone=\sum_n|n\rangle\langle n|$, where $n=0$ corresponds to $|0\rangle=|\psi(h_1,h_2)\rangle$ such $H_{\rm pr}(h_1,h_2)=\sum_{n\neq 0}|n\rangle\langle n|$. This relation implies that all other energies are equal to one. The Hamiltonian is not local in terms of the original Landau orbitals. Nevertheless, it has the same quasihole state as an eigenstate of a physically motivated local Hamiltonian, and since the evolution is adiabatic, all other eigenstates, which distinguish this Hamiltonian from a local one, do not participate in the evolution. Once we make the quasihole wave functions continuous, the parent Hamiltonian by construction inherits the continuous dependence on $h_1$, lending itself well to an adiabatic process with time-dependent $h_1$. Furthermore, the gap between the quasihole state and all other eigenstates is constant regardless of the position of the quasihole.

To generate adiabatic braiding, we slowly move the quasihole coordinate $h_1$ on the contours by integrating the Schr\"{o}dinger equation 
\beg\label{schEq}
i\hbar\frac{\partial \ket{\psi(h_1,h_2)}}{\partial t}=\hat{H}_{pr}(h_1(t),h_2(t))\ket{\psi(h_1,h_2)}.
\en
Each of the four legs of the trajectory is traversed over a time $T$, with the entire process completed in time $4T$. We have found that for $L_x\sim 10-20\ell_B$, timescales of $T=500-1000$ are sufficient for maintaining adiabaticity with $>0.999$ overlap between the final and the initial states. As discussed in Appendix.~\ref{app:unphys}, for contours ${\cal C}_3$ and ${\cal C}_4$, we take a time $T/2$ to bring the $y$ component of $h_1(t)$ to $y_2$, the $y$ component of $h_2$ for the stationary quasihole. This choice allows us to make the wave function continuous.
The time dependence of $h_1$ is summarized below:
\begin{center}
{
\begin{tabular}{ccc}
\hline\hline 
Contour  & Time & $h_1(t)$\\ 
\hline 
$\mathcal{C}_1$  & ~~~$0<t<T$~~~ & $[L_x (1-t/T),y_{t}]$\\ 
$\mathcal{C}_3$  & $T<t<3T/2$ & $[0,y_t-2(y_t-y_2)(t-T)/T]$\\ 
$\mathcal{C}_3$  & $3T/2<t<2T$ & $[0,y_2-(y_2-y_b)(2t-3T)/T]$\\ 
$\mathcal{C}_2$ &  ~~~$2T<t<3T$~~~ & $[L_x(t-2T)/T,y_b]$ \\ 
$\mathcal{C}_4$ & $3T<t<7T/2$ & $[L_x,y_b+2(y_2-y_b)(t-3T)/T]$ \\ 
$\mathcal{C}_4$ & $7T/2<t<4T$ & $[L_x,y_2+(y_t-y_2)(2t-7T)/T]$ \\ 
\hline \hline
\end{tabular} }\end{center}
During the entire process, the position of the second quasihole is time-independent: $h_2(t)=[L_x/2,y_2]$.

The results of these numerical simulations are shown in Fig. \ref{Fig:FullPath}. The Berry phase is shown as a function of the flux $\Phi$ (modulo $6\pi$) for various sets of parameters. It exhibits excellent agreement with Eq. \eqref{analytic}.
As mentioned earlier, for paths $\mathcal{C}_3$ and $\mathcal{C}_4$, a continuous interpolation of the topological sector is used as discussed in Appendix ~\ref{app:unphys}. The results are obtained by directly solving the Schr\"odinger equation as an ordinary differential equation.

\section{Quantum Algorithm and Berry Phase measurements}\label{sec:algorithm}
\subsection{Efficient basis of physical qubits}
To extract the Berry phase from adiabatic braiding on small, noisy quantum devices, we seek basis states that efficiently encode the states participating in the dynamics and maximally utilize the quantum resources. In the Laughlin case, the basis states are labeled by $q_1$ and $q_2$ in Eq. (\ref{hole2}). The unitary $\hat{S}$ is independent of $h_1$ and $h_2$ \cite{Siedel13} and therefore the states $\hat{S}(L_x,0)\ket{c,q_1,q_2}$ form an orthonormal basis:
\beg
\nonumber
\langle c,q_1,q_2|\hat{S}^\dagger(L_x,0)\hat{S}(L_x,0)|c',q'_1,q'_2\rangle=\delta_{c,c'}\delta_{q_1,q'_1}\delta_{q_2,q'_2}.
\en

The number of relevant basis states $\hat{S}(L_x,0)\ket{c,q_1,q_2}$ is quadratic in the number of orbitals. An efficient quantum algorithm should utilize all $2^M$ states of $M$ physical qubits. Thus, we map these states to the bitstrings of the physical qubits and make the gate decomposition local in terms of the physical qubits. This is done by choosing an ordering of the $\hat{S}(L_x,0)\ket{c,q_1,q_2}$ states for a fixed $c$ and assigning a bitstring to each state such that the bitstring is the binary representation of the position of the state in the ordered list. We can further improve the basis's efficiency by truncating the Gaussian amplitudes' tails and keeping a subset of all allowed $q_1$ and $q_2$.

Further simplification is possible by noting that for $c=0$, placing $\mathcal{C}_1$ at a multiple of $3\kappa$ will contribute to the Berry phase by a multiple of $2\pi$, which can be seen from the expression for the Berry phase. The contribution of $\mathcal{C}_1$ to the Berry phase is given by
\beg\label{numerical}
\begin{split}
\gamma_1&=-i\oint_{\mathcal{C}_1} d\vec{l}. \bra{\psi_c(h_1,h_2)} \nabla_{h_1} \ket{\psi_c(h_1,h_2)}\\
&=-2\pi{\cal N}_2^2 \sum_{q_1<q_2}q_1e^{\frac{1}{3}(h_{2y}-\kappa b_{c,q_2})^2}e^{\frac{1}{3}(h_{1y}-\kappa b_{c+1,q_1})^2}.
\end{split}
\en
The above expression is equal to $-2\pi \langle q_1\rangle$. Note that the integrand becomes independent of $h_{1x}$, and the effect of the integration is a factor of $L_x$. Although the summation is constrained to $q_1<q_2$, due to the Gaussian decay of the probability amplitudes, most of the contribution comes from $q_1$ and $q_2$ near the peak of the two Gaussians. Thus relaxing the constraint leads to minimal error. In the absence of the constraint, the sum factorizes, and the average of $q_1$ corresponds to the peak of one Gaussian as $\langle q_1\rangle={h_{1y}\over3\kappa}-1$, which is an integer when ${h_{1y}\over3\kappa}$ is an integer, leading to a trivial contribution to the Berry phase. Thus, our simulations can be carried out for $\mathcal{C}_2$ only, given that $\mathcal{C}_3$ and $\mathcal{C}_4$ cancel out. To control the enclosed flux, we change the position $y_b$ of $\mathcal{C}_2$, while keeping $y_t$ fixed to an integer multiple of $3\kappa$.

Since we only implement the braiding around $\mathcal{C}_2$, with a stationary $h_2$, we can pin $q_2$ to a single site $q_2^*$ corresponding to the peak of the Gaussian without losing any significant topological feature.
Assuming the suitable truncated range of $q_1$ starts at $q_1^{\min}$, the explicit identification of the basis states is given below
\beg
\begin{split}
\hat{S}(L_x,0)\ket{0,q_1^{\min},q_2^*}&=\ket{0...00},\\
\hat{S}(L_x,0)\ket{0,q_1^{\min}+1,q_2^*}&=\ket{0...01},\\
\hat{S}(L_x,0)\ket{0,q_1^{\min}+2,q_2^*}&=\ket{0...10}.\\
&\vdots \\
\hat{S}(L_x,0)\ket{0,q_1^{\min}+2^{M}-1,q_2^*}&=\ket{1...11}.\\
\end{split}
\label{armin_basis}
\en
Our truncation criterion for $\mathcal{C}_2$ is that $\ket{0,q_1,q_2}$ if $|\phi_0(h_1,q_1)||\phi_1(h_2,q_2)|>10^{-6}$. The widths of the Gaussian amplitudes are controlled by $\kappa$ and hence $L_x$. Therefore, the larger $L_x$, the more physical qubits are required for emulating the process.

\subsection{Ground state preparation}\label{subsec:GS} 
Since we have compressed the number of physical qubits exponentially, we use generic (exponentially expensive) state preparation schemes to prepare the initial state that remains polynomial in the number of electrons.
A general state for $M$ qubits can be written as $\ket{\Psi}=\sum_{i=0}^{2^M-1}a_i\ket{i}$, where $N=2^M$ is the total number of basis states. At the start of  $\mathcal{C}_2$, the amplitudes of the states are related to the wave function as $a_0=\mathcal{N}\phi_{0,q_1^{\min}}(h_1 )\phi_{1,q_2^*}(h_2)$, $a_1=\mathcal{N}\phi_{0,q_1^{\min}+1}(h_1)\phi_{1,q_2^*}(h_2)$. The general algorithm of Ref. \cite{GroundCircuit} then requires  $O(NM^2)$ {one- and two-qubit quantum gates} and utilizes the following single-qubit unitary:
\beg\label{GuiUnitary}
U_0(\alpha)=\begin{bmatrix}
\cos(\alpha) & \sin(\alpha) \\
\sin(\alpha) & -\cos(\alpha)
\end{bmatrix}.
\en
This unitary can be written as a special case of the unitary
\beg
U_3(\theta,\phi,\lambda)=\begin{bmatrix}
    \cos(\frac{\theta}{2}) & -e^{i\lambda}\sin(\frac{\theta}{2})\\
     e^{i\phi}\sin(\frac{\theta}{2}) & e^{i(\phi+\lambda)}\cos(\frac{\theta}{2})
\end{bmatrix}
\en
for $\theta=2\alpha$, $\phi=0$ and $\lambda=\pi$. In the first step, we apply the unitary in the Eq, (\ref{GuiUnitary}) for an angle $\alpha_1=\tan^{-1}\bigg(\sqrt{{\sum_{i_2,,..i_n}|a_{0i_2i_3..i_n}|^2}/{\sum_{i_2,..i_n}|a_{1i_2i_3..i_n}|^2}}\bigg)$ to create the superposition $\sqrt{\sum_{i_2,i_3..i_n}|a_{0i_2 i_3..i_n}|^2}\ket{00..0}+\sqrt{\sum_{i_2,i_3,..i_n}|a_{1 i_2 i_3..i_n}|^2}\ket{10..0}$. In the next step, we create the appropriate superposition of $\ket{00..00}$, $\ket{01..00}$, $\ket{10..00}$ and $\ket{11..00}$ by applying two controlled rotations by angle $\alpha_{20}=\tan^{-1}\bigg(\sqrt{{\sum_{i_3,..i_n}|a_{01i_3i_4..i_n}|^2}/{\sum_{i_3,..i_n}|a_{00i_3i_4..i_n}|^2}}\bigg)$ and $\alpha_{21}=\tan^{-1}\bigg(\sqrt{{\sum_{i_3,..i_n}|a_{11i_3i_4..i_n}|^2}/{\sum_{i_3,..i_n}|a_{10i_3i_4..i_n}|^2}}\bigg)$ to the second qubit, controlled by the first qubit; the $\alpha_{20}$ and $\alpha_{21}$ rotations are applied when the first qubit is in state 0 and 1 state respectively. In the next step, four doubly controlled rotations are performed on the third qubit in the same manner. This sequence continues until the last step, where we perform $2^{M-1}$ controlled rotations controlled by all ${M-1}$ first qubits. In the last step, the rotations can involve the complex amplitudes of the basis states:
\beg\label{GuiCmplxUn}
U_{\alpha_n i_1 i_2..i_{n-1}}=\begin{bmatrix}
\frac{A_0}{\sqrt{|A_0|^2+|A_1|^2}} & \frac{A_1}{\sqrt{|A_0|^2+|A_1|^2}} \\
\frac{A_1^*}{\sqrt{|A_0|^2+|A_1|^2}} & -\frac{A_0}{\sqrt{|A_0|^2+|A_1|^2}} 
\end{bmatrix},
\en
where $A_0=a_{i_1 i_2 ..i_{n-1}0}$ and  $A_1=a_{i_1 i_2 ..i_{n-1}1}$ and $a_{i_1 i_2 ..i_{n-1}1}$. An example of the circuit decomposition for one segment of the trajectory is presented in Appendix~\ref{appn:gr_circuit}.

\subsection{Braiding unitary circuit}
As discussed above, the unitary operators that emulate the braiding process are obtained from constructing a parent Hamiltonian labeled by the quasihole position, changing the position by several increments along a cyclic trajectory, and determining the individual unitaries in the sequence. Then, the unitaries corresponding to different segments of the quasihole trajectory are decomposed in terms of local quantum circuits. 

%Gate decomposition in terms of physical qubits will be obtained by using the machine-learning approaches of The Berkeley Quantum Synthesis Toolkit (BQSKit).
Gate decomposition in terms of physical qubits will be obtained by using the approaches \cite{qiskit.org,PRAiso}. Once we create these unitary circuits, we can directly apply the sequence of all these circuits to an initial two-quasihole eigenstate prepared by the circuit discussed above. Each unitary moves the quasihole by a certain distance emulating its physical motion. The product of all these unitaries, which we refer to as $U_{\rm braid}$, brings the state back to the initial eigenstate up to the overall Berry phase. An example of the circuit decomposition for one segment of the trajectory is shown in Appendix~\ref{appn:circuit}.

\subsection{Measurement of the Berry phase}

\begin{figure}[]
\includegraphics[width=\columnwidth]{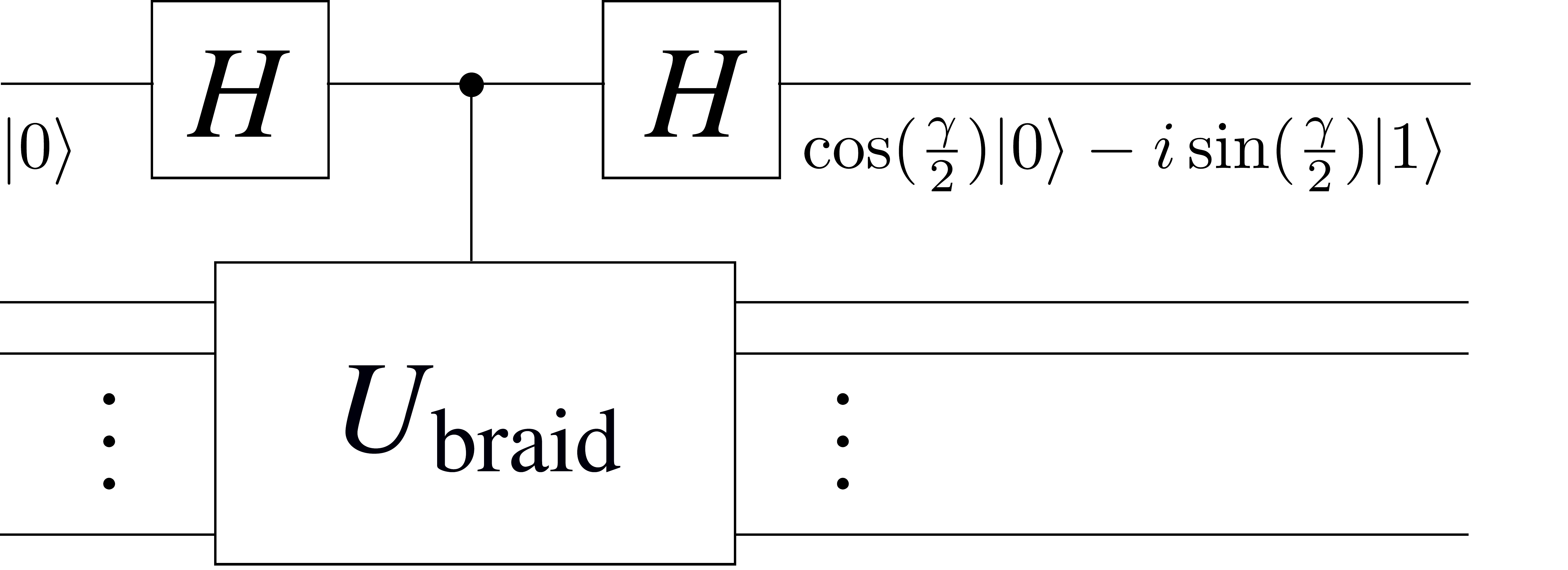}
	\caption{Using an ancilla controlling the braiding unitary to extract the Berry phase $\phi$ from measuring the probabilities of 0 and 1 for the ancilla qubit.\label{fig:meas} }
\end{figure}

We need to extract the Berry phase by measurements at the end of the process. As we can only read out probabilities, which are independent of the overall phase of the wave function, measurements on the qubits representing the quantum Hall state are insufficient. However, using an ancilla qubit allows the extraction of the exchange statistics. We implement a more complex circuit shown in Fig.~\ref{fig:meas}, where the ancilla controls the entire $U_{\rm braid}$ obtained by applying a sequence of unitaries for different segments of the trajectory on the initial quasihole state. The construction of this controlled unitary is straightforward once $U_{\rm braid}$ is decomposed into one and two-qubit gates, as each gate in the circuit for $U_{\rm braid}$  is promoted to a control gate, controlled by the ancilla. After this promotion, the circuit for each segment is then optimized using 
%the machine-learning approaches of 
the Berkeley Quantum Synthesis Toolkit (BQSKit) \cite{bqskit}. An example of the ancilla-appended optimized circuit is presented in Appendix~\ref{appn:circuit}.

A comment is in order regarding this approach. Because the net effect of the whole product of the unitaries on the initial state will be a phase factor, it is crucial to apply $U_{\rm braid}$ as a sequence of adiabatic transformations without any circuit-level simplification that goes across the individual unitaries for different trajectory segments. While we individually optimize the unitary (controlled by the ancilla) for each segment, we perform no optimization on the product of these unitaries. By keeping the individual segment unitaries intact in the sequence, we transform the state through the intermediate positions of the quasihole, producing a genuine adiabatic braiding process in the quantum hardware.

\section{Quantum hardware results}\label{sec:ibm}
\begin{figure}[!h]
    \includegraphics[width=7cm]{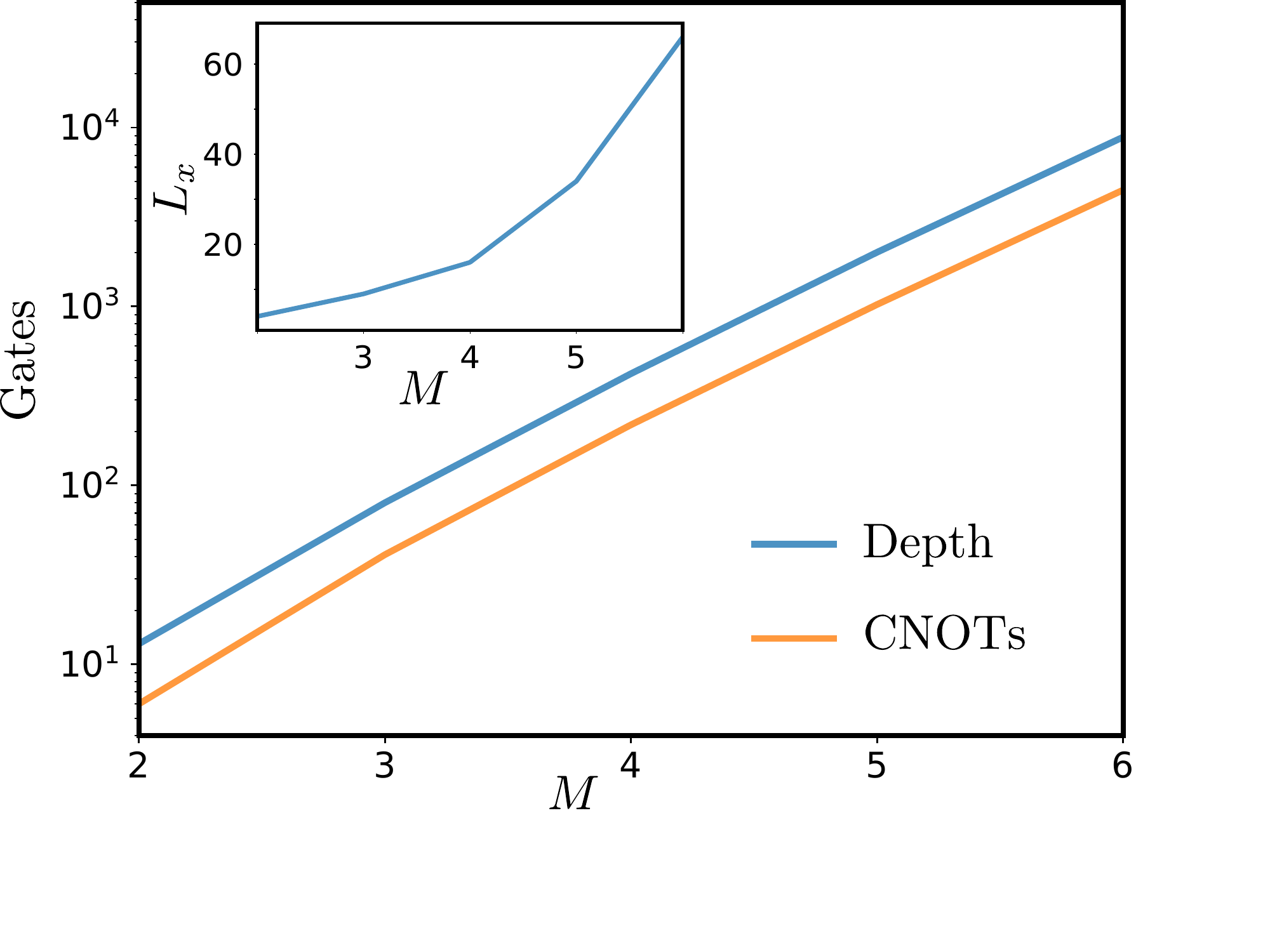}
	%\centering
	\caption{Circuit depth and the number of CNOTs as a function of the number of physical qubits $M$. The inset shows the circumference $L_x$ that can be simulated for a given $M$. }
 	\label{Fig:CircuitDepth}
\end{figure}
\subsection{Results on exchange statistics}

 To generate quantum circuits for each adiabatic path, we input segment unitaries with dimensions $2^M\times 2^M$ obtained from numerics on $\mathcal{C}_2$ into the routines provided by Qiskit \cite{qiskit.org,PRAiso}. The circuit depth of a single segment is independent of the number of total segments in an adiabatic trajectory. The depth of each segment for a 10-segment path is shown in Fig. \ref{Fig:CircuitDepth} as a function of $M$. After the unitary circuit of each segment is obtained, we append the ancilla qubit to each segment's circuit, by promoting every gate in the circuit to a control gate. We then optimize the resulting circuit for each segment and run them on IBM devices.
\begin{figure*}[ht!]
    \centering
    \includegraphics[width=15cm]{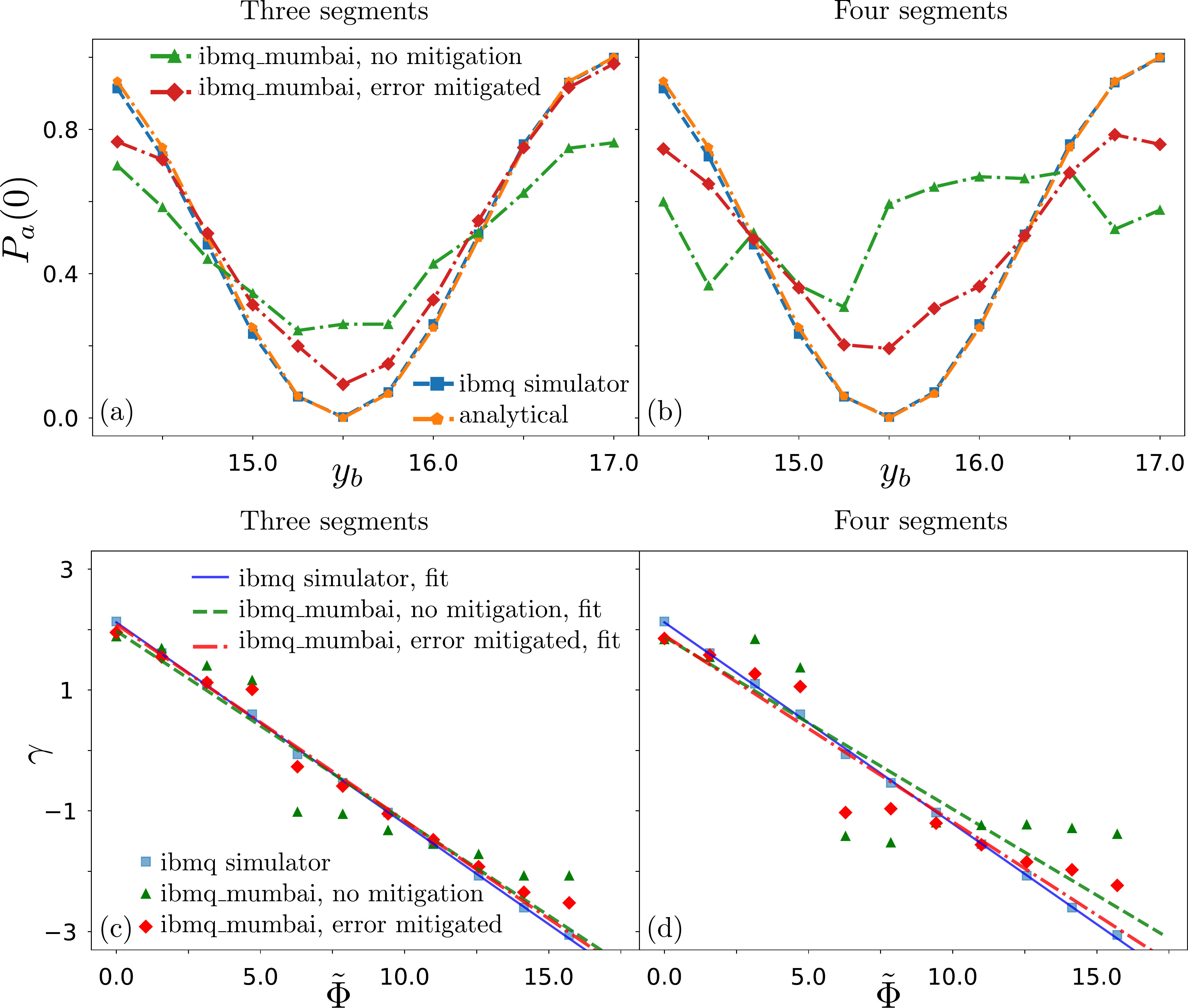}
    \caption{Braiding FQH quasiholes on quantum hardware.
    (a)-(b) The probability $P_a(0)$ of measuring the ancilla in the 0 state after braiding.
    (c)-(d) The resulting flux-dependent Berry phase, as a function of $\tilde{\Phi}=\Phi~{\rm mod}~6\pi$.
    The results were obtained with two physical qubits plus one ancilla. We compare results from analytical calculations (`analytical'), IBM's noiseless Aer simulator, and quantum hardware without and with error mitigation. (a) and (c) panels are the results for the braiding path with 3 segments, while (b) and (d) are for 4 segments. This data was collected between December 6, 2022 and January 25, 2023 on \texttt{ibmq$\_$mumbai}. The median CNOT and readout errors were 0.9\% and 2\%, respectively, that varied between the qubits. There were small variations of median CNOT and readout errors between these dates. 
    In the case of 3 (4) segments, the slopes are -0.32 (-0.31) and the intercept are 2.09 (1.90).}
    \label{fig:Main}
\end{figure*} 
The ancilla-appended circuits are then optimized with BQSKit to reduce the gate count and appended to the ground-state preparation circuit to complete the path of $\mathcal{C}_2$.

Here we use two qubits for the unitary dynamics and one ancilla qubit for the measurements. We move quasihole $h_1$ along the path $\mathcal{C}_2$ in 3 or 4 steps, each corresponding to the quasihole movement by length $L_x/3$ or $L_x/4$. Our state are labelled as $\ket{0,2,p}\to\ket{00}$, $\ket{0,3,p}\to\ket{01}$, $\ket{0,4,p}\to\ket{10}$ and $\ket{0,5,p}\to\ket{11}$. To obtain the Berry phase as a function of the flux, we move $\mathcal{C}_2$ by changing $y_b=14.25\kappa$ to $17\kappa$ in small steps. For each value of $y_b$, we implement the initial state circuit on IBM devices using the method described in the Sec. \ref{subsec:GS}. We then apply the braiding unitary that is controlled with the ancilla qubit. Once the $h_1$ makes a complete circle around $\mathcal{C}_2$, we measure the ancilla in the computational basis and record the probability of the ancilla qubit being in the state $|0\rangle$. This probability is expected to be equal to $[1+\cos(\gamma)]/2$. This probability is shown in Fig. \ref{fig:Main}(a)-(b) for three- and four-segment trajectories, respectively, with and without error mitigation techniques. From the ancilla probabilities, we extract the Berry phase and compute its slope and the intercept as a function of the flux. We find the slope and the intercept from the error-mitigated results, -0.32 and 2.09 (-0.31 and 1.90), for three (four) segments to be in agreement with the analytically expected results of $1/3$ and $2\pi/3\approx 2.09$. The Berry phase is shown in Fig. \ref{fig:Main}(c)-(d) for the three- and four-segment trajectories, respectively.

\subsection{Error mitigation}
We mitigate errors on IBM's superconducting processors with a composite strategy comprised of error suppression and error mitigation methods adapted from, e.g., Ref.~\cite{Ferris2022SimTutorial}. 
To suppress coherent errors, we twirl \cite{Wallman2017Twirling} each circuit 20 times, such that the total number of twirled circuit times shots per circuit is conserved at 8,000, and insert evenly spaced XY4 dynamical decoupling pulses \cite{Ezzell2022DDSurvey} in idle periods to reduce coherent error. 

Resulting noisy bitstrings from each shot are mitigated by first correcting readout error with the so-called `M3' method \cite{Nation2021M3}, where the number of calibration circuits scales favorably linearly with the number of qubits. Then, the bitstrings probabilities are post-selected given physical constraints. In particular, it is known \textit{a priori} that braiding simply applies a global phase to the original state. Therefore, the noisy bitstring probabilities of the system qubits marginalized over the ancilla qubit should be the same for both the initial and braided states, so we filter the latter's bitstring probabilities to match the former's. Notably, this procedure does not leverage classical simulations of the braided system, as the bitstring probabilities for both the initial and braided states are measured on the quantum simulator. Finally, we apply zero-noise extrapolation \cite{Li2017ZNE, Temme2017PECandZNE}, where we linearly extrapolate the ancilla probability to the zero-noise value using the expectations with 8,000 shots at noise factors of 1 (corresponding to the original circuit) and 3 (where each gate in the transpiled circuit is locally folded once using the Qiskit Prototype \cite{Rivero2022ZNE}). The total shots overhead for the entire composite error mitigation strategy is thus only two times the original number of shots.

\section{Conclusions} \label{sec:conclusions}

In this paper, we use a quasi-1D model of $\nu=1/3$ FQH state as a framework to develop an efficient quantum algorithm for braiding quasiholes on a quantum computer. We execute this algorithm on a small, noisy IBM quantum computer and directly measure the fractional statistical angle after Hamiltonian evolution corresponding to adiabatic braiding. To do so, we directly construct a parent Hamiltonian from the two-quasihole wave function, map orthonormal states that are superpositions of the occupation basis states of the Landau orbitals to bitstrings of physical qubits, and break the process into several segments. Analogous to phase estimation, we then control the entire adiabatic braiding circuit with an ancilla qubit, which allows us to extract the Berry phase from probability measurements on the ancilla.  Finally, we apply a suite of quantum error mitigation techniques to improve the hardware results significantly. To simplify the process, we also chose a particular braiding path, where two legs cancel by construction.

Moving forward, we expect the development of more powerful quantum devices would make it possible to access more general trajectories and use the physical qubits in one-to-one correspondence with the local orbitals. Paired with further advances in the error mitigation techniques leveraged in the present work, such devices could then be used to study different problems in the dynamics of correlated electron systems, such as the topological and geometric excitations of more exotic quantum phases like non-Abelian fractional Hall states \cite{nayak2008non} through suitable adiabatic and quench processes, as well as emergent phenomena in topological Floquet systems.

\section*{Acknowledgements} We thank Claudio Chamon, Sriram Ganeshan, Layla Hormozi, and Zelatko Papic for helpful discussions. We acknowledge the use of IBM Quantum services. We also thank the Brookhaven National Laboratory for providing access to IBM devices. A.K. was
supported by NSF Grants No. DMR-2037996 and No. DMR-2038028. P.G. acknowledges support from NSF Awards No. DMR-2037996 and DMR-2130544. A.R. acknowledges support from NSF Award No. DMR-2038028 and  Award No. DMR-1945395. A.R. and P.G. also acknowledge the Aspen Center for Physics, which is supported by NSF Grant No. PHY-1607611. A.R. acknowledges Boston University Physics Department for hospitality.

\appendix
\section{Interpolation of the topological sector for ${\cal C}_{3,4}$ contours}\label{app:unphys}
Contributions to the Berry phase in the present work come from contours $\mathcal{C}_1$ and $\mathcal{C}_2$, where one quasihole is significantly away from the other quasihole and $\kappa(q_1-q_2)\gg1$. For contours $\mathcal{C}_3$ and $\mathcal{C}_4$, the $y$ coordinate of one quasihole crosses that of the nonmoving hole. In general, $\kappa(q_1-q_2)\gg1$ does not hold as $q_1\sim q_2$ configurations are not suppressed during the crossing. Although the contribution from $\mathcal{C}_3$ and $\mathcal{C}_4$ cancel out for our particular geometry, to investigate the full cyclic process, we need a continuous Hamiltonian. Thus we utilize a scheme to circumvent the abrupt change of the topological sector and phase of the wave function. We give the expression of the wave-function amplitudes that only differ substantially from Eq. \eqref{eq:factor} in the regime where the validity of Eq. \eqref{eq:factor} breaks down but avoids any discontinuity arising from topological sector change.

We focus on contour ${\cal C}_3$, which is traversed over time $T<t<2T$. A similar construction is used for contour ${\cal C}_4$. Let $t'=t-T$ be the time parameter along $\mathcal{C}_3$. For $0<t'<T/2$, the quasihole moves vertically from $y_t$ to $y_2$. Eq. \eqref{eq:factor} in this case is equal to $\phi_{0,q_1,q_2}(h_1,h_2)\approx \mathcal{N}_2 \phi_{0,q_1}(h_2)\phi_{1,q_2}(h_1)$. Similarly for $T/2<t'<T$, Eq. \eqref{eq:factor} gives $\phi_{0,q_1,q_2}(h_1,h_2)\approx \mathcal{N}_2 \phi_{0,q_1}(h_1)\phi_{1,q_2}(h_2)$, where the quasihole moves from $y_2$ to $y_b$. We first make the magnitude of the wave function continuous by changing the topological sector continuously so that $|\phi_{0,q_1,q_2}(h_1,h_2)|$ is given by
\begin{center}
{
\begin{tabular}{ccc}
\hline\hline  
$|\phi_{0,q_1,q_2}(h_1,h_2)|$ & $t'$ \\ 
\hline 
$|\mathcal{N}_2 \phi_{t'/T,q_1}[h_2(t)]\phi_{1-t'/T,q_2}[h_1(t)]|$ & ~~~$0<t'<T/2$~~~ \\ 
$|\mathcal{N}_2 \phi_{1-t'/T,q_1}[h_1(t)]\phi_{t'/T,q_2}[h_2(t)]|$  & $T/2<t'<T$ \\ 
\hline \hline
\end{tabular} }\end{center}
We can further make the phase continuous. For simplicity, we take $h_{1x}=0$ and $h_{2x}=L_x/2$. Then the phase of the wavefunction switches from $e^{-iq_1\kappa L_x/2}=e^{i\pi q_1}$ to $e^{-iq_2\kappa L_x/2}=e^{i\pi q_2}$ at the crossing. We can use a phase $e^{i [\pi-\alpha(t')]q_1+i\alpha(t')q_2}$, where $\alpha(t')$ is respectively close to 0 and $\pi$ before and after crossing and continuously interpolates between these two values in close vicinity of the crossing.

\section{Ground state circuit}\label{appn:gr_circuit}

\begin{figure}[t]
\includegraphics[width=\columnwidth]{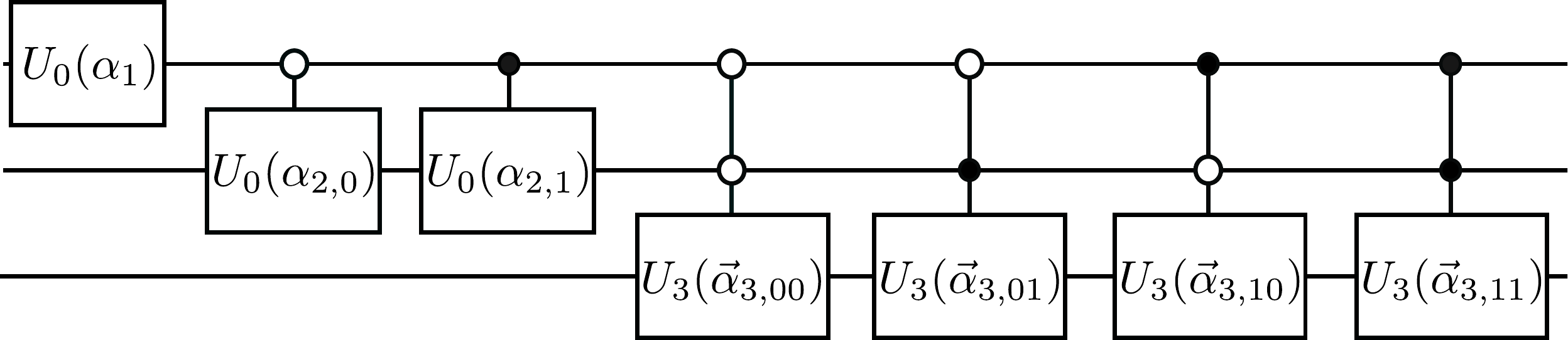}
	\caption{A three-qubit ground state circuit. }\label{fig:GS} 
\end{figure}
\begin{figure}[b]
    \includegraphics[width=8cm]{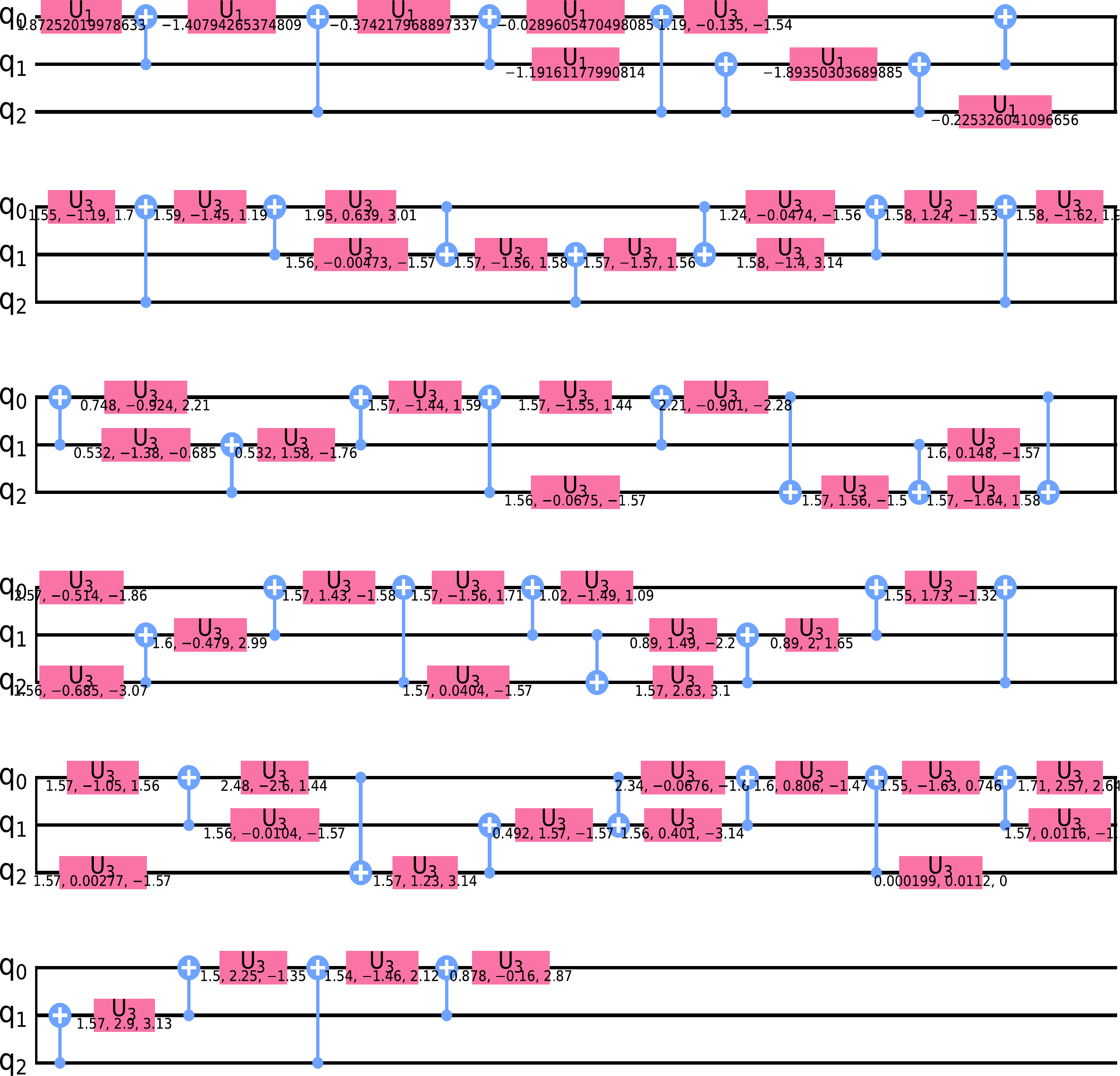}
	%\centering
	\caption{A 3-qubit unitary segment.}
    \label{fig:qiskitUn}
\end{figure}
For a state with $L_x=10\ell_B$ and $h_1=(0,14.5\kappa)$, with amplitudes
\begin{eqnarray*}
a_{000}&=&2.69\times 10^{-5},\quad a_{001}=-2.07\times 10^{-3},\quad\\ a_{010}  &=&4.86\times 10^{-2},\quad a_{011} =-3.50\times 10^{-1},\\
a_{100} &=&7.71\times 10^{-1},\quad
a_{101}=-5.19\times 10^{-1},\\
a_{110}&=&1.07\times 10^{-1},\quad
a_{111}= -6.75\times 10^{-3},
\end{eqnarray*}
we find
$\alpha_1=1.21$ in $U_0({\alpha_1})$ and two controlled $U_0$ rotations by $\alpha_{2,0}=1.565$ and $\alpha_{2,1}=0.115$. For the last stage, we apply four doubly controlled $U_3$ gates with angles $\vec\alpha_{3,00}=(-2\times 1.558,0,0)$, $\vec\alpha_{3,01}=(-2\times 1.433,0,0)$, $\vec\alpha_{3,10}=(-2\times 0.593,0,0)$ and $\vec\alpha_{3,11}=(-2\times 0.063,0,0)$, where we denote the three arguments of the $U_3$ unitary by $\vec{\alpha}=(\theta,\phi,\lambda)$. We are able to write the unitary of Eq.~\eqref{GuiCmplxUn} as a $U_3$ operator because our state is real for $h_1=(0,h_{1y})$ at the beginning of the adiabatic process.
\section{Circuit details}\label{appn:circuit}
The unitaries for braiding a hole around $\mathcal{C}_2$ are obtained by numerically solving the adiabatic Schr\"{o}dinger equation on a classical computer. We then choose our qubit basis according to Eq. (\ref{armin_basis}) and obtain the corresponding quantum circuits through Qiskit functions \cite{qiskit.org, PRAiso} that deterministically decompose unitaries into one-qubit gates $U_3$ and  $U_1(\lambda)=\begin{bmatrix}
    1 & 0\\
     0 & e^{i\lambda}
\end{bmatrix}$ and two-qubit {\small CNOT} gates, as shown in Fig \ref{fig:qiskitUn} for a 3-qubit system where $L_x=9\ell_B$ and $\mathcal{C}_2$ is placed at $12.5 \kappa$. The circuit corresponds to a single segment.

\begin{figure}[]
    \includegraphics[width=8cm]{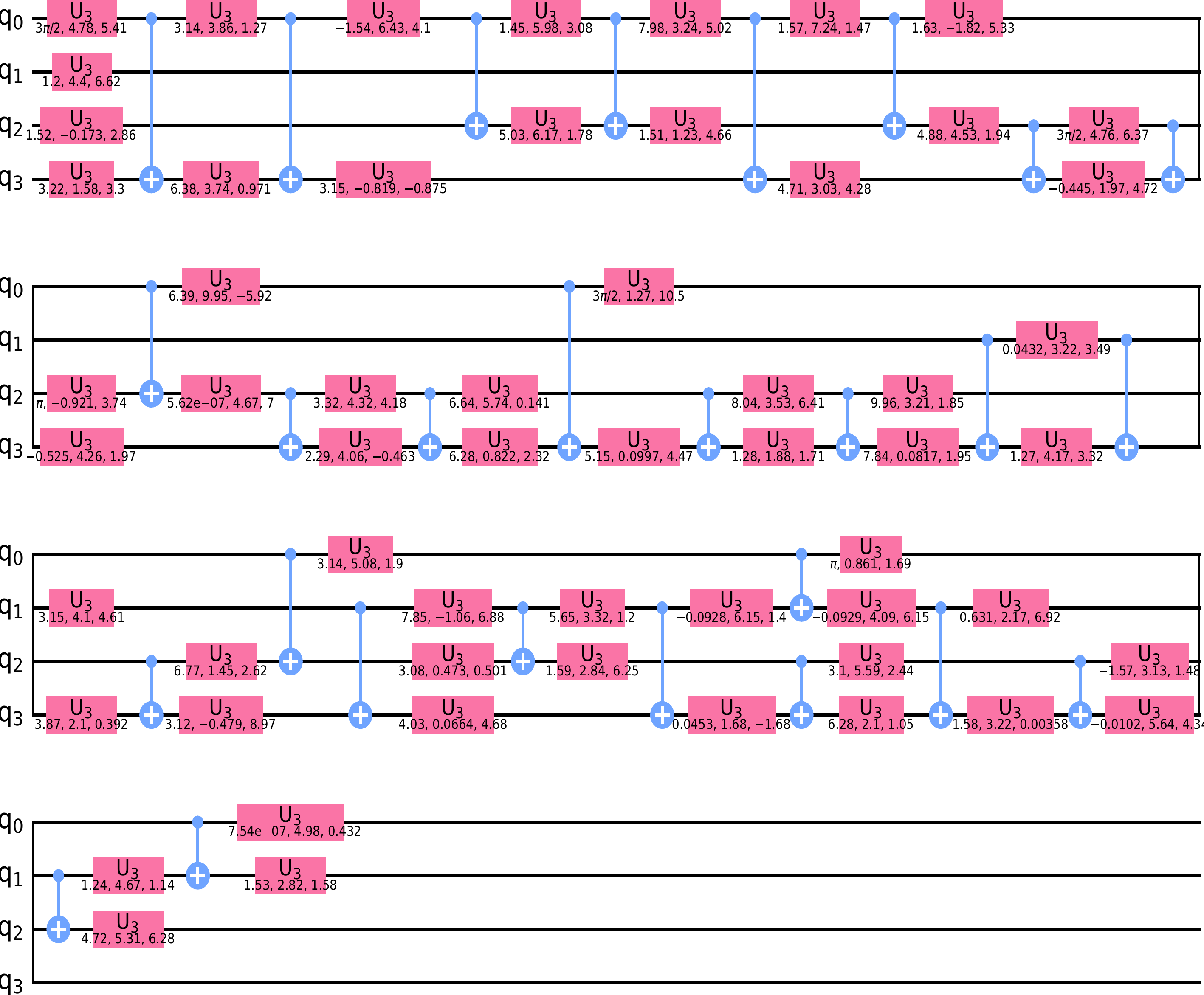}
	%\centering
	\caption{A 3-qubit unitary segment when controlled by ancilla qubit after circuit optimization.}
    \label{fig:bqUn}
\end{figure}

We then append the ancilla qubit to this circuit, which increases the depth of the circuit by changing the rotation gate to an ancilla-controlled rotation and the {\small CNOT} gate to a Toffoli gate. To reduce the gate count, we then re-transpile the ancilla-appended circuits in BQSKIT. For example, the unitary segment of Fig. \ref{fig:qiskitUn} with an appended ancilla and that has been re-transpiled is given in Fig. \ref{fig:bqUn}. This process is repeated for each segment to obtain a complete unitary, as shown in Fig. \ref{fig:meas}.

\bibliography{braid}
\end{document}